\documentstyle[pre,aps,multicol,epsf]{revtex}

\tighten

\begin{document}
\draft

\title{ Collective Dynamics of Josephson Vortices in 
Intrinsic Josephson Junctions:  
Exploration of In-phase Locked Superradiant Vortex Flow States 
}  

\author{$^{a}$ M. Machida}

\address{ 
Center for Promotion of Computational Science and Engineering,
\\ 
 Japan Atomic Energy Research Institute, 2-2-54 Nakameguro, Meguro-ku
Tokyo 153, Japan 
\\
$^{a}$ CREST, Japan Science and Technology Corporation(JST), Japan
}

\author{ T.Koyama }
\address{ Institute for Materials Research, Tohoku University, Katahira
2-1-1, Aoba-ku, Sendai 980-77, Japan}

\author{ A.Tanaka }
\address{ Teikyo University of Science, Uenohara 2525, Yamanashi, Japan } 

\author{ M.Tachiki }
\address{
 National Research Institute for Metals, Sengen 1-2-1, Tsukuba, Ibaraki 305, Japan}

\date{\today}

\maketitle

\begin{abstract}

In order to clarify the ``superradiant'' conditions for the 
moving Josephson vortices to excite 
in-phase AC electromagnetic fields over all junctions, 
we perform large scale simulations of realistic dimensions for intrinsic 
Josephson junctions under the layer parallel magnetic field. 
Three clear step-like 
structures in the I-V curve are observed above a certain high 
field ( $H > 1T$ in the present simulations ), 
 at which 
we find structural transitions in the moving flux-line lattice. 
The Josephson vortex flow states are accordingly classified 
into four regions ( region I $\sim $  IV with  
increasing current ), 
in each of which the power spectrum for the electric 
field oscillations at the sample edge are measured and  
typical snapshots for Josephson vortex configurations are displayed. 
Among the four regions, especially in the region III,  
 an in-phase 
rectangular vortex lattice flow state 
emerges and 
the power spectrum shows remarkably sharp peak structure, i.e., 
superradiant state.  
Comparison of the simulation results with an eigenmode analysis for 
the transverse propagating Josephson plasma oscillations reveals that 
the resonances between Josephson vortex flow states and some of the eigenmodes 
are responsible for 
the clear flux lattice structural transitions. 
Furthermore, the theoretical analysis clarifies that 
the width of the superradiant state region in the I-V characteristics 
enlarges with decreasing both the superconducting and insulating layer thickness. 


\end{abstract}

\vskip2pc  



\begin{multicols}{2} \narrowtext

\section{Introduction}



Intrinsic Josephson effects (IJE's) in 
highly anisotropic high temperature 
superconductors (HTSC's) such as Bi$_2$Sr$_2$CaCu$_2$O$_8$ 
is a subject currently under intensive investigation \cite{Kleiner1}.
This is because 
single crystals of these materials naturally form
a stack of many atomic-scale Josephson junctions, i.e., intrinsic Josephson
junctions (IJJ's) and therefore 
strong coherence 
between junctions 
are expected. 
Especially, if in-phase oscillation over all junctions is realized
it will show superradiant properties such as
remarkable increase of output power and decrease of 
linewidth [2-7], 
and its realization will 
give a great impact on technological applications to 
optical communications in the near future.   
Therefore, in this paper, we explore such superradiant conditions in
Josephson vortex flow states in IJJ's.  

In the case of the single long Josephson 
junctions (SLJJ's),  fluxon dynamics coupled with transverse propagating
Josephson plasma oscillation have already been extensively 
investigated \cite{Barone},\cite{Abrikosov},\cite{Zhang1}, and
SLJJ's have been employed as useful flux flow oscillators.
In SLJJ's, the fluxons are aligned like one-dimensional chain along the 
junction plane and  
unidirectionally driven under the bias-current. 
This sliding motion of the fluxon chain excites 
the transverse propagating Josepshson plasma oscillations  
inside the junction, and 
when their moving 
speed matches the Josephson plasma mode velocity, 
I-V characterisitcs 
show a step like resonant structure, where the 
electromagnetic (EM) wave radiation power becomes maximal 
\cite{Zhang1},\cite{See1}. 

On the other hand, 
IJJ's have a longitudinal propagating 
Josephson plasma mode
along the stacked direction 
and its dispersion relation 
are shown to be very different from that of
the transverse mode, i.e., the longitudinal one
is almost flat while the transverse one is very dispersive. 
Thus, IJJ's can have many propagating modes with different dispersions as 
mixtures between transverse
and longitudinal modes, and IJJ systems with finite 
stacking height (employed in experiments) have 
the standing waves  
for those mixed modes along the stacked direction 
\cite{Kleiner2},\cite{Sakai1}.  
Accordingly, the I-V characterisitics and the 
properties of EM radiation in IJJ's are clearly expected to exhibit a
rich variety in contrast to SLJJ's. 
Actually, so far, the fluxon dynamics in IJJ's 
have been investigated in several papers[2-7],\cite{Artemenko1} to understand their 
dynamics and related EM radiation, and 
complex resonant behaviors with some special modes have been reported. 
However, the full understanding for the 
behavior of the fluxons and Josephson plasma oscillations has not been 
achieved. 
Especially, the clear in-phase superradiant conditions in IJJ's 
have not yet been established, and 
some questions still now remain concerning the 
stability of the superradiant flow state and its emergent region 
with respect to experimental variables as the magnetic field and 
the transport current. 
Therefore, in this paper, we report results of the first direct large scale 
simulations for the fluxon dyanamics in IJJ's using the 
coupled Sine-Gordon equation   
and clarify the superradiant conditions.
Furthermore, we pay attention to 
all dynamical flux flow states observed in various current ranges in 
the I-V characteristics and 
reveal that the I-V characteristics show 
three clear step-like structures corresponding to  
flux lattice structural transitions.
Thus, we distinguish flux flow states into four 
regions at these three step-like structures, and 
display the typical flux flow lattice configuration  
in each of the regions. The names region I $\sim$ IV 
( in increasing order of the current ) will be assigned to 
the four regions.
In order to identify the natures of the EM radiations in the four regions,  
the electric field oscillations at the sample edge, which  
characterize the properties of the emitted electromagnetic wave, are 
measured and their typical power spectrum in each region is given.

Here, let us briefly describe the characters of 
the flux flow states observed 
in the four regions. The notable points are remarkable differences in 
the flux lattice configurations and power spectra for EM field oscillations 
in those four regions. 
In region I, fluxons flow randomly  without forming a
clear lattice structure. This state is a steady state when the applied current 
is low enough, and the electric field oscillations at the edge exhibit
a broad-band power spectrum. 
In region II, wavy aligned flux flow states appear. From this 
region, the vortex flow states
begin to show resonances with some of the transverse propagating modes.
The vortex alignments develop along the stacked direction, and 
the power spectrum 
has peak structures at a characteristic frequency and its higher harmonics.
In region III, a perfect in-phase flux flow states emerges. 
The vortex alignment becomes perfect along the c-axis 
and fluxons form a rectangular lattice. The line-width of the power spectrum 
is very sharp and the power at the peak position is also very large, i.e., 
the superradiant state. 
In region IV, triangular like lattice flux flow states 
are observed. Its Josephson frequencies due to the flux flow voltages 
are larger than the frequencies of any plasma modes.
Consequently, it is expected that 
the transverse plasma modes does not affect
the formation of flux lattice structures and therefore 
energetic favorable triantular like lattice flow states appear.
In order to understand the origin of the transitons 
between these four flux flow states, we 
perform an eigenmode analysis on the employed coupled Sine-Gordon equation.
As a result, we are able to identify 
that the resonances between three modes and the
flux flow states leads to 
three clear step-like structures in the I-V characteristics.
Thus, from this theoretical analysis, we can  predict the width of 
the four regions, especially, the superradiant state (region III),  
in the I-V characteristics.
Also, the analysis clarifies that 
the superradiant state region enlarges with increasing  
a parameter ratio, $\frac{ {\lambda_{ab}}^2 }{s D}$, where $s$ and $D$ 
are, respectively, superconducting and insulating layer thickness 
and $\lambda_{ab}$ is  the penetration depth in the ab-plane direction.
This result clearly indicates that  
IJJ's can have remarkably wide superradinat region in I-V characterisitics.

The outline of this paper is as follows. 
The model equation employed 
is briefly introduced in section II, and the numerical simulation results 
are shown in section III. Section IV is devoted to eigen mode analysis and 
discussions. 

\section{Model Equation} 

We first explain our model. We perform the simulations under a 
configuration in which the transport current flows in the direction 
perpendicular to the junctions ($\parallel$ $z$-axis) and the magnetic 
field is applied along the junctions ($\parallel$ $y$-axis) as shown in Fig.1. 

\begin{figure}
\centerline{\epsfxsize=9.4cm \epsfysize=7.0cm \epsfbox{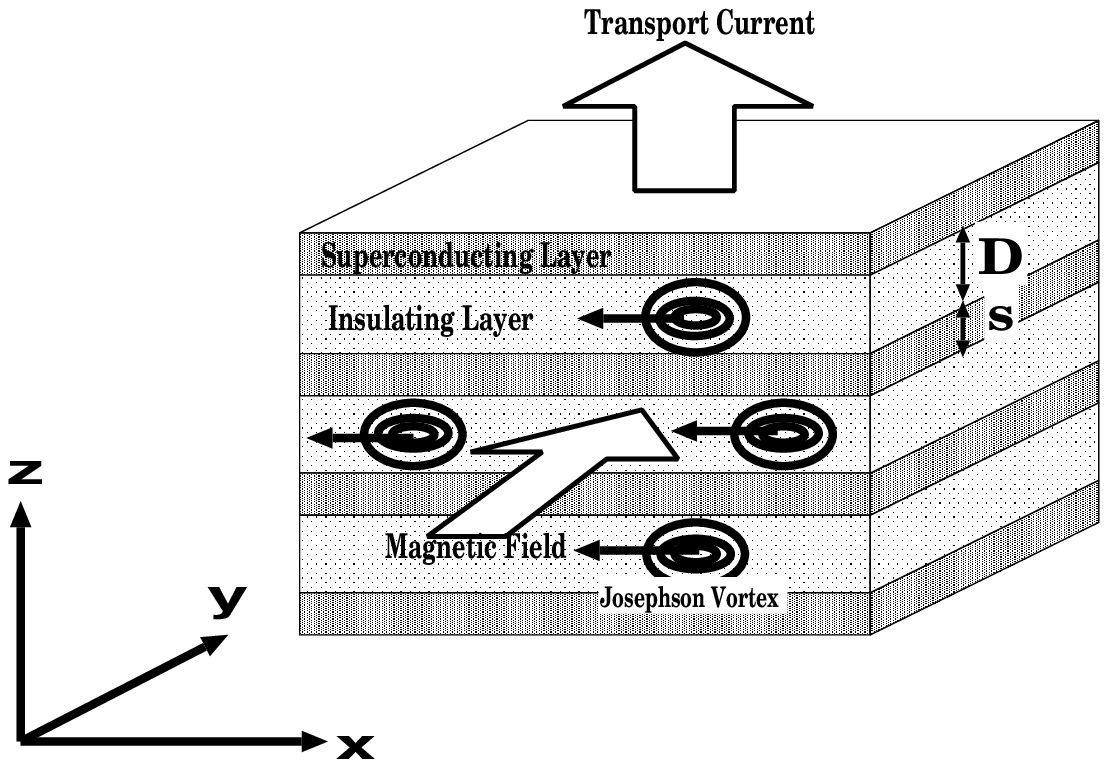}} 
\end{figure}
FIG.1~ A schetch of the intrinsic Josephson junction. The magnetic 
field and the transport current are applied along the junction plane (y-axis) 
and the stacked directions (z-axis), respectively. The Josephson vortices
move in the -x direction. 
\bigskip

\noindent The Josephson vortices move parallel to the $-x$-axis in this 
configuration. Assuming that the superconducting layers have an 
atomic-scale thickness, we neglect the variation of physical 
quantities along the $z$-axis inside the superconducting layers. 
Then, the vector potential is 
defined only inside the insulating layers as 
$A_{\ell+1,\ell}^z(x,t)(\equiv\int_{\ell}^{\ell+1} dz A_z (x,t) )$, 
where $\ell$ stands for the layer index,  
and the electromangtic fields $E_{\ell+1,\ell}^z(x,t)$ and
$B_{\ell+1,\ell}^y(x,t)$ are, respectively given by 
$E_{\ell+1,\ell}^z(x,t)=-\frac{1}{cD} 
\frac{\partial A_{\ell+1,\ell}^z }{\partial t}
-\frac{\varphi_{\ell+1} - \varphi_{\ell} }{D}$, and
$B_{\ell+1,\ell}^y(x,t) = \frac{A_{\ell+1}^x - A_{\ell}^x}{D}
- \frac{1}{D}\frac{\partial A_{\ell+1,\ell}^z }{\partial x} $.
We assume Maxwell's equations in the present system as follows, 
\begin{eqnarray} 
\frac{ \partial {B^y}_{\ell+1,\ell} }{ \partial x} 
= \frac{4\pi}{c} {j^{z}}_{\ell+1,\ell} 
+ \frac{ \epsilon_c}{c} \frac{ \partial {E^z}_{\ell+1,\ell} }{\partial t}, \\
 {E^z}_{\ell+1,\ell} - {E^z}_{\ell,\ell-1} = \frac{4 \pi s}{\epsilon_c} 
\rho_{\ell},   
\end{eqnarray}
where $\rho_{\ell}$ and $\epsilon_c$ are, respectively, the charge density 
at $\ell$-th superconducting layer and the dielectric constant of the insulating
layers. 
The current, ${j^{z}}_{\ell+1,\ell}$ in eq.(1) is given 
by a sum of Josephson and quasiparticle currents as $j_c sin P_{\ell+1,\ell} 
+ \sigma {E^z}_{\ell+1,\ell}$, with $j_c$ and $\sigma $
being the critical current density and conductivity of
the quasiparticles, respectively. 
$P_{\ell+1,\ell}$ is the gauge invariant phase difference defined as
$P_{\ell+1,\ell}= \theta_{\ell+1} - \theta_{\ell} - 
\frac{2\pi}{\phi_0} A_{\ell+1,\ell}$ with $\theta_{\ell}$ being the phase of
the superconducting order parameter of $\ell$-th layer.
The time and spatial derivatives for $P_{\ell+1,\ell}$ 
yield the generalized relations,  

\begin{eqnarray}
\frac{\partial P_{\ell+1,\ell}}{\partial t} = -{4 \pi {\mu}^2}
\frac{2 \pi c}{\phi_0} ( \rho_{\ell+1} - \rho_{\ell} )
+ \frac{2 \pi c D}{\phi_0}{{E^z}_{\ell+1,\ell}},\\ 
\frac{\partial P_{\ell+1,\ell} }{ \partial x} 
= \frac{4\pi {\lambda_{ab}}^2 }{c} \frac{2\pi}{\phi_0} ( {{j^x}_{\ell+1}}
- {{j^x}_{\ell}} ) + \frac{2 \pi D}{ \phi_0} {{B^y}_{\ell+1,\ell}},
\end{eqnarray}
where $\lambda_{ab}$ are the penetration depth associated with 
the current ${j^x}_{\ell}$ in the ab-plane and $\mu$ is the Debye length 
\cite{Tachiki1} \cite{Mac1}. 
In deriving eq.(3) and (4), we employ the equation 
$\rho_{\ell} = -\frac{1}{4 \pi \mu^2} (\frac{\phi_0}{2\pi c} \partial_t 
\theta_{\ell} + \varphi_{\ell}) $, which is analogous to the London equation and 
the London equation for the in-plane current,  
${j^x}_{\ell} = -\frac{c}{4\pi{\lambda_{ab}}^2} 
( \frac{\phi_0}{2\pi} \nabla_x \theta_{\ell} - {A^x}_{\ell} ) $.
In the case where $j_{\ell+1}^x-j_\ell^x \not=0$ under an external 
magnetic field the first term on the right hand side of eq.(3) gives 
only small contribution of $O(\mu^2/\lambda_{ab}^2)$ to the dynamics 
of $P_{\ell+1,\ell}$, which means that the incomplete screening effect 
for charges in the superconducting layers, namely the coupling between 
junctions due to the charge neutrality breaking 
effect \cite{Tachiki1},\cite{Mac1} may be neglected in the 
presence of an external magnetic field. Then, in this approximation 
the equation for $P_{\ell+1,\ell}$ is derived from eqs.(1)-(4) as 
follows  \cite{Tachiki1}, \cite{Bulaevskii1},\cite{Shafranjuk} 

\begin{eqnarray}  
\frac{ {\partial}^2 P_{\ell+1,\ell}}{\partial {t'}^2 } 
& = &    \frac{{\lambda_{ab}}^2}{s D} 
( \frac{\partial^2 P_{\ell+2,\ell+1} }{\partial {t'}^2} 
+ \frac{ \partial^2 P_{\ell,\ell-1}}{\partial {t'}^2} 
-2 \frac{ \partial^2 P_{\ell+1,\ell} }{\partial {t'}^2} ) \nonumber \\
&+& \frac{ {\lambda_{ab}}^2 }{ s D } ( sin P_{\ell+2,\ell+1} + 
sin P_{\ell,\ell-1}
- 2 sin P_{\ell+1,\ell} ) \nonumber \\
& + &\beta \frac{ {\lambda_{ab}}^2 }{ s D } 
( \frac{ {\partial} P_{\ell+2,\ell+1}}{\partial {t'} } +
 \frac{ {\partial} P_{\ell,\ell-1}}{\partial {t'} }
-2 \frac{ {\partial} P_{\ell+1,\ell}}{\partial {t'} } ) \nonumber \\
& + &\frac{{\partial}^2 P_{\ell+1,\ell} }{ \partial {x'}^2 } 
-  sin P_{\ell+1,\ell} - 
\beta \frac{ {\partial} P_{\ell+1,\ell}}{\partial {t'} }, 
\end{eqnarray}
where $t^\prime=\omega_pt$ and $x^\prime=\lambda_cx$ with $\omega_p$ 
and $\lambda_c$ being, respectively, the Josephson plasma frequency 
($\omega_p={c\over\sqrt{\epsilon}\lambda_c}$) and the penetration depth 
in the $c$-axis direction. The parameter $\beta(\equiv {4\pi\sigma \lambda_c
\over\sqrt{\epsilon}c}$) is related the McCumber parameter $\beta_c$ as 
$\beta_c=1/\beta^2$. It is noted that eq.(5) gives a flat dispersion 
relation for the longitudinal mode along the $c$-axis while it leads to 
the dispersive one along the junctions.

\section{Results of Numerical Experiments and Discussions}

  In this paper, we numerically solve eq.(5) for a finite size IJJ composed 
of 20 junctions in which the $ac$ cross section ( the $xy$-plane in Fig.1) 
has 2-dimensional (2D) rectangular shape in the presence of an applied 
field ($\parallel y$-axis) and a transport current ($\parallel z$-axis). 
The transport current and the magnetic field which includes the self-field 
are introduced through the boundary conditions at the top and bottom layers 
and the edges of all the layers. To perform the simulations we divide the 
system into 400 meshes along the $x$-direction. Each mesh is assumed to be 
$0.1\lambda_{ab}$ wide along the $x$-direction. Then the system used in 
the simulations has the area of $10\mu m\times 20$ junctions if we assume 
$\lambda_c=125\mu m$ and $\gamma=500$ for the anisotropic parameter, which 
are typical parameter values in Bi$_2$Sr$_2$CaCu$_2$O$_8$. The value of 
the material parameter $\beta$ is chosen as $\beta=0.1(\beta_c=100)$, which 
is also a typical value in Bi$_2$Sr$_2$CaCu$_2$O$_8$. The results shown 
in this paper are qualitatively size-independent. We confirmed this by 
performing the simulations for system sizes up to $30\mu m\times 80$ 
junctions in real scale. 

  Let us now present the results of numerical simulations. The applied 
magnetic field used in the following calculations is fixed to a value, 
$2T$. The results for lower fields ($<1T$), which are qualitatively 
different, will be published elsewhere. Figure 2(a) shows the $I-V$ 
characteristics of IJJ. As seen in this figure, three clear steps appear 
at the voltage values indicated by arrows in the $I-V$ curve. To 
understand the origin of the steps, we plot the time and spatial 
averages of the Poynting vector, $<P_{-x}>=<{c\over 4\pi S}
\int dx\sum_\ell E_{\ell+1,\ell}^xB_{\ell+1,\ell}^y>$, and the Josephson 
current, $<J_z>=<{1\over S}\int dx\sum_\ell\sin P_{\ell+1,\ell}>$ in 
Fig.2 (b) and (c), where $<>$ stands for the time average and $S$ is 
the area of the computational region. Note that $<P_{-x}>$ represents 
the energy flow of the transverse EM field and $<J_z>$ gives a measure 
of the viscosity against the driven flux-flow. It is seen that the 
anomalous structures also appear in these averages, i.e., the step-like 
structure in $<P_{-x}>$ and the peaks in $<J_z>$, at the same voltage 
values as in the $I-V$ characteristics. Such anomalies are commonly 
observed in our simulations in the case where the magnetic field is 
greater than 1T. Thus, it is found that the vortex dynamics are 
directly coupled with the transverse EM field and the vortex viscosity
drastically changes at the step like structures. 
Especially, it is noted that the viscosity of the Josephson vortices 
takes maximum values at the steps. From this fact one may expect that 
the collective vortex dynamics changes at the steps. In the following, we, 
therefore, separate the voltage region into four from I to IV at the steps 
as shown in Fig.2(a) and investigate the collective properties of each 
flux-flow state separately.


\begin{figure}
\centerline{\epsfxsize=9.4cm \epsfysize=15.0cm \epsfbox{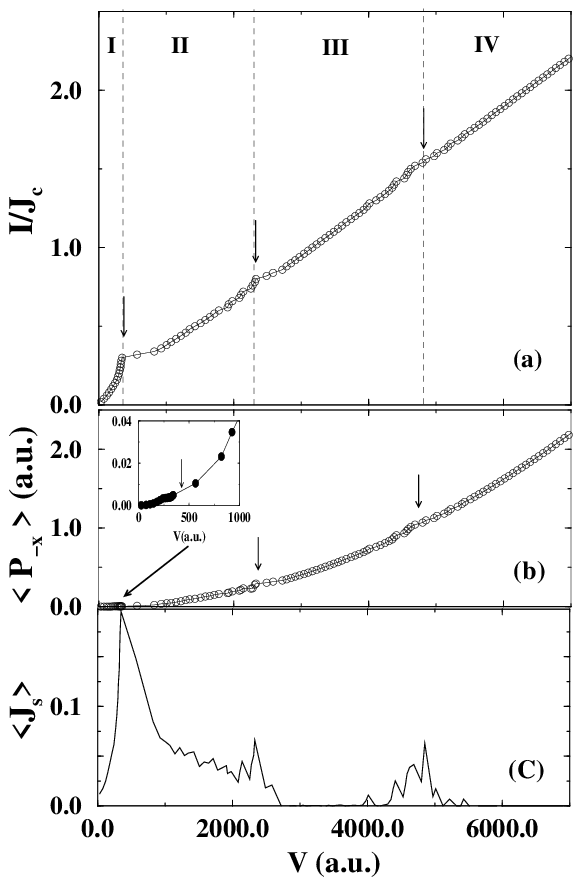}} 
\end{figure}
FIG.2~ (a) The I-V characteristics. (b) The averaged 
poynting vector in x-direction $< P_{-x} >$ vs. V. The inset of (b) gives 
the enlarged view of the region indicated by the inclined arrow.
(c) The averaged Josephson current vs. V.
\bigskip

\begin{figure}
\centerline{\epsfxsize=8.4cm \epsfysize=11.0cm \epsfbox{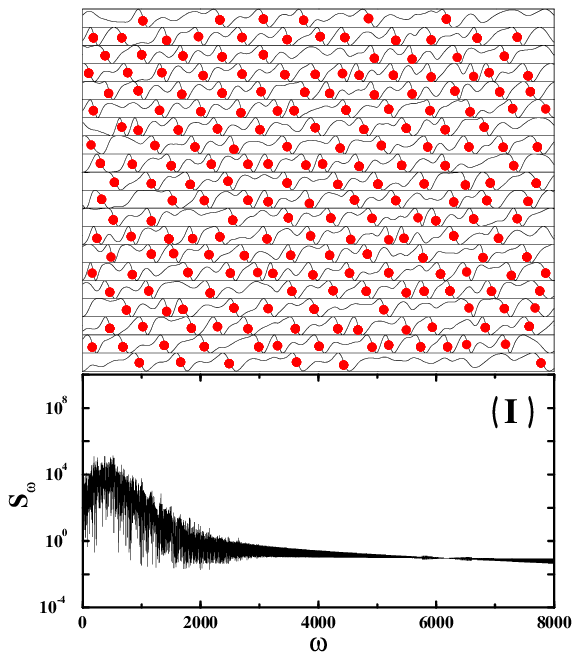}} 
\end{figure}
\vskip -2.2cm
\begin{figure}
\centerline{\epsfxsize=8.4cm \epsfysize=11.0cm \epsfbox{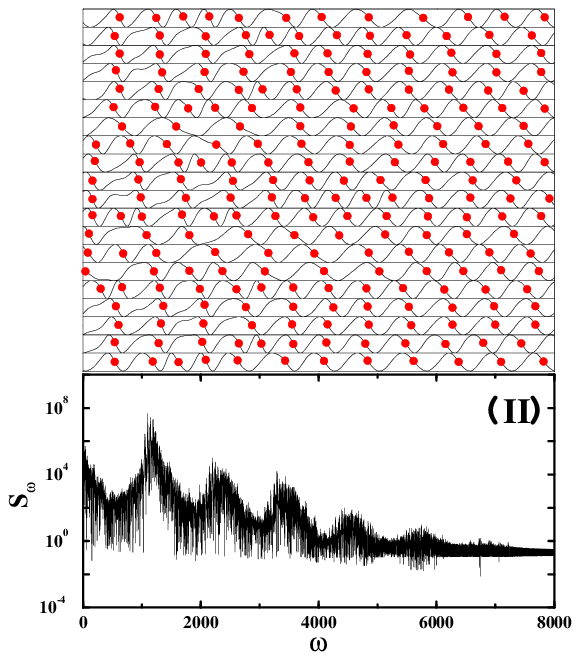}} 
\end{figure}
\vskip -1.4cm
FIG.3, The snapshot of the vortex center positions along with 
$sinP_{\ell+1,\ell}$ distributions in all junction sites and the
power spectrum of the electric field at the left edge
in the regions from I to IV: (I) $I/J_c = 0.20$, (II) $I/J_c = 0.52$.
Note that the Josephson plasma frequency at the zero temperature,
$\omega_p = \frac{c}{\sqrt{\epsilon_c} \lambda_c (0) }$, corresponds to  
$ \omega=130 $ in the figures for the power spectrum.    
\bigskip


  Let us study the flux-flow state in region I first. In Fig.3(I) we show 
a snapshot of the distribution of the Josephson current, $\sin P_{\ell+1,
\ell}$, and the centers of the 
Josephson vortices ( indicated by dots ) together with 
the power spectrum of the electric field at the left edge of the sample, 
$S_\omega\equiv {2\pi\over T}|{1\over 2 \pi}\int_0^T\sum_\ell E_{\ell+1,
\ell}(0,t){\rm e}^{i\omega t}{\rm d}t|^2$. As seen in this figure, the 
vortices are distributed almost randomly and the power spectrum shows 
$\omega^{-2}$ dependence above some frequency, which indicates the 
irregular motion of vortices. We confirmed this chaotic behavior of moving 
vortices by monitoring directly the time development of the vortex 
positions too. From these results one may conclude that the highly viscous 
flux-flow state is realized in region I. The fluxons are easily pinned 
by any type of imhomogeneities, e.g., crystal imperfections, sample boundary 
etc. in this low driving-current regime. In the present system without 
imperfections the sample boundaries including the top and bottom junctions 
work as effective pinning centers, which is seen from the fact that the 
number of vortices in the top and bottom junctions is much smaller than 
that in other intermediate junctions. It is also noticed that the power 
spectrum forms a broad band in the low frequency region. Thus, the chaotic 
viscous flux motion generates a broad-band oscillating EM field in region 
I. The intensity of the emitted EM field increases with increasing 
voltage values in this region.


    Let us next turn to the region II. As seen in the snapshot given in 
Fig.3(II), a wavy-chain-like vortex distribution is stabilized in this 
region. This result indicates that a correlation between vortices is 
developed in region II. Accordingly some characteristic frequency and 
its higher harmonics appear inside the broad band spectrum of the EM field. 
These frequencies are related to the quasi-periodicity of the moving vortex 
lattice. It is also noted that the frequency of the highest peak seen in 
the figure is about eight times larger than the Josephson plasma frequency 
at zero temperature, $\omega_p(={c\over\sqrt{\epsilon}\lambda_c})$. This 
frequency corresponds to about 4THz in the case of $\lambda_c=125\mu m$ 
and $\epsilon_c=25$.


When the voltage is further increased and reaches the value at the 
boundary between the regions II and III, the moving vortices form a 
rectangular lattice as shown in Fig.3(III). This regular distribution of the 
vortices is stable and maintained throughout the region III. It is seen 
that sharp peaks appear in the power spectrum of the EM field. The 
frequencies giving the peaks are determined by the flux flow velocity and the 
period of the lattice. Note that the phase differences of all the junctions 
make perfect in-phase motion in this state, that is, 
{\it the superradiant state is realized in the region III}. In this 
superradiant state one may expect the strong radiation of the coherent 
EM field from the junctions. The frequencies of the emitted EM field will 
be tunable in a wide range by the applied magnetic field or the voltage 
since the region III is broad as seen in the $I-V$ curves. 
 

\begin{figure}
\centerline{\epsfxsize=8.4cm \epsfysize=11.0cm \epsfbox{Fig3ct.eps} }
\end{figure}
\vskip -2.2cm
\begin{figure}
\centerline{\epsfxsize=8.4cm \epsfysize=11.0cm \epsfbox{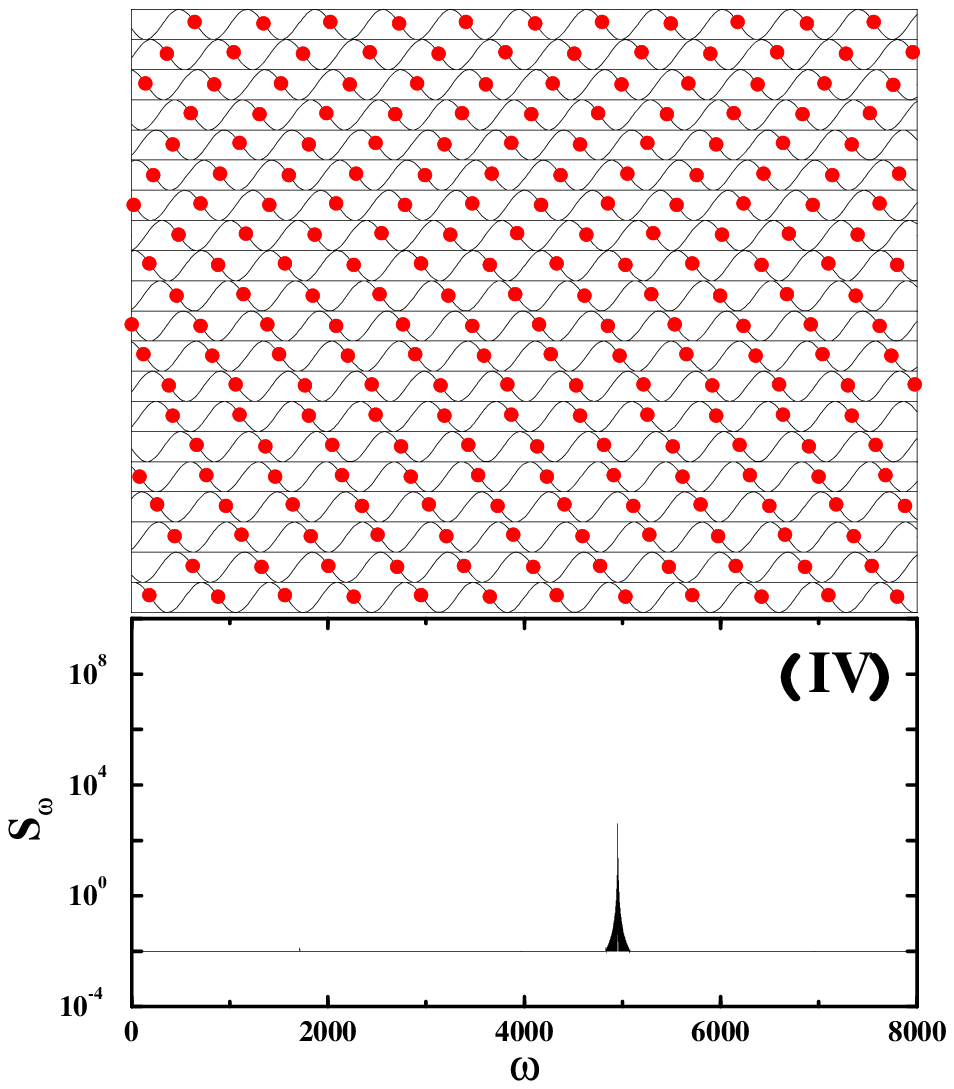}} 
\end{figure}
\vskip -1.4cm
FIG.3 (III) $I/J_c = 1.16$, (IV) $I/J_c = 1.90$.
\bigskip

The rectangular lattice of the moving vortices becomes unstable at the boundary 
between the regions III and IV. The moving vortex lattice appearing in the 
region IV is a modulated triangular one as depicted in Fig.3(IV). The 
power spectrum given in Fig.3(IV) indicates that the intensity of the EM 
field generated in this state is greatly reduced from that in the region 
III. This result is easily understood in the following way. The oscillatory 
components of the tunneling current on two consecutive junctions cannot 
be in-phase in a triangular-like lattice. As a result, the EM field 
generated at the junction edges on consecutive junctions interfere 
destructively with each other and, therefore, the intensity of the power 
spectrum drastically decreases in this region.

\section{ Eigen Mode Analysis and Discussion }

To understand the origin of the transitions in the moving vortex state 
we perform a linear analysis in the following. The linearized equation 
obtained from eq.(5) in the case of $\beta=0$ yields the eigenvalues, 
$\omega_n(k_x)=\{1+k_x^2/[1+{2\lambda_{ab}^2\over sD}
(1-\cos{n\pi\over N})]\}^{1/2}$, and the eigenfunctions, 
$P_{\ell+1,\ell}(x^\prime,t^\prime)={\rm e}^{ik_xx^\prime}
{\rm e}^{i\omega_nt^\prime}f_{\ell+1,\ell}^{(n)}$ with 
$f_{\ell+1,\ell}^{(n)}=g_n{\rm e}^{i\pi\ell n/N}$ for the system with $N$ 
junctions, where $n$ specifies the eigenmode, i.e., $n=1,\cdots, N-1$. 
These eigenmodes may be interpreted to be the transverse Josephson plasma 
modes in finite IJJ composed of $N$ junctions without vortices. In the inset 
of Fig.4(a) we plot the spatial variation of ${\rm Re}(f_{\ell+1,\ell}^{(n)})$ 
for four values (n=0,1,2,and 19) of $n$ in the case of $N=20$ , which 
represents the 
standing waves along the $c$-axis.
In the presence of an external magnetic 
field, $H$, and a DC voltage, $V$, eq.(5) has an approximate solution 
corresponding to the phase-locked flux-flow state as  
$P_{\ell+1,\ell}(x^\prime,t^\prime)=H^\prime x^\prime - V^\prime t^\prime 
+ h_{\ell+1,\ell}(x^\prime,t^\prime)$, where $H^\prime={2\pi D\lambda_c
\over\phi_0}H$ and $V^\prime={2\pi c\over\phi_0\omega_p}V$. Here, the 
function $h_{\ell+1,\ell}(x^\prime,t^\prime)$ is assumed to be an small 
oscillatory component. In this case, $h_{\ell+1,\ell}(x^\prime,t^\prime)$ 
satisfies approximately the equation, 
$[(1-{\lambda_{ab}^2\over sD}\Delta^{(2)})\partial_t^2-\partial_x^2]
h_{\ell+1,\ell}=-\sin (H^\prime x^\prime - V^\prime t^\prime)$, 
where $\Delta^{(2)}$ is defined by $\Delta^{(2)}h_{\ell+1,\ell} 
=  h_{\ell+2,\ell+1} +  h_{\ell,\ell-1} -2 h_{\ell+1,\ell}$ 
\cite{Kleiner2},\cite{Barone}.
This equation has solutions showing resonances at $\omega \sim H^\prime/
\{[1+{2\lambda_{ab}^2\over sD}(1-\cos{n\pi\over N})]\}^{1/2}$. 
This resonant behavior originates as a result that the Josephson 
oscillations induced by the moving flux-line lattice match the transverse 
Josephson plasma frequencies. 
Note that the resonant frequencies coincide 
with the eigen-frequencies $\omega_n(k_x)$ for $k_x=H^\prime\gg 1$. 
Figure 4(a) shows the matching voltages ($V^\prime=\omega_n(k_x=H^\prime)$) 
as a function of $H$ in the case of $N=20$. In Fig.4(b), we superpose the 
vertical lines indicating the matching voltages at $k_x=H(=2T)$ on the 
$I-V$ curve. As seen in this figure, the voltages matching with the modes 
A ($n=0$) and B ($n=1$) coincide with the 3rd and 2nd steps and the lowest 
matching voltage is also located close to the 1st step in the $I-V$ curves. 
These analyses indicate that the structural transitions in the moving 
flux-line lattice are caused by the matching of the phase oscillations 
arising from the vortex motion with that of the eigen-modes of the system. 

\begin{figure}
\centerline{\epsfxsize=7.4cm \epsfysize=11.5cm \epsfbox{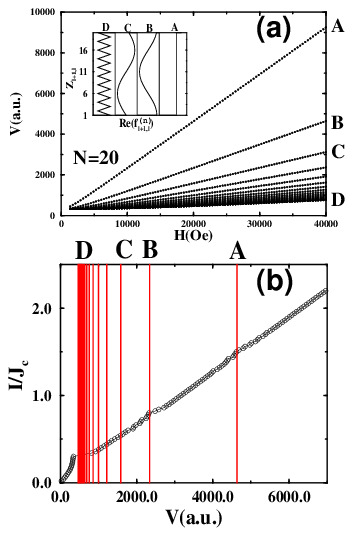}} 
\end{figure}
FIG.4, (a)The magnetic field dependence of the matching voltage 
between the flux flow states and all transverse eigenmodes. The inset
shows the variations along c-axis of $Re({f^{(n)}}_{\ell+1,\ell})$ of 
A$(n=0)$, B$(n=1)$, C$(n=2)$ and 
D$(n=19)$ modes, where ${f^{(n)}}_{\ell+1,\ell}=g_n e^{i \pi \ell n/N}$.     
(b) The I-V characterisitcs same as Fig.1(a). The vertical lines indicate
the matching voltage of all eigenmodes, and A, B, C, and D correspond to  
those shown in Fig.4(a), respectively   

\bigskip

\noindent Here, it is noted that not all the matching voltages cause clear 
step-like anomalies 
in the $I-V$ characteristics. However, we have directly observed the structural 
change, for example, near the voltage value at C by monitoring 
the vortex configuration. 
Since the $I-V$ characteristics is given by a sum of the contribution from 
all the junctions, it is understood that it is not so sensitive for a 
slight structural change in the moving flux-line lattice, that is, in a case 
that the current distribution is not largely altered at the transition, a clear 
anomalous structure will not appear in the $I-V$ characteristics. 
We also notice that the matching with the modes, A and B, obtained in the 
above linear analyses gives the precise phase boundaries of the region 
III. This fact indicates that the above analysis is sufficiently valid in
 this region, that is, the oscillatory part $h_{\ell+1,\ell}$ is 
only a small perturbation and the phase-locked in-phase 
flux-flow state, which is the moving rectangular-lattice state, is very stable. 
The width of the region III is given by $\Delta V=
\omega_A(H)-\omega_B(H)\sim [1-{\sqrt{sD}\over \sqrt{2}\lambda_{ab}}
{1\over\sqrt{1-\cos(\pi/N)}}]H$. Since ${\sqrt{sD}\over \sqrt{2}\lambda_{ab}}$ 
is very small in high-$T_c$superconductors, the superradiant state is expected 
to be very wide in IJJ's.


Hechtfischer et al. observed a broad-band emission in the flux-flow state 
of Bi$_2$Sr$_2$CaCu$_2$O$_8$ single crystals, whose power spectrum takes 
a maximum around the voltage at which the step-like structure appears in 
the $I-V$ characteristics \cite{Hechtfischer1}. They claim that 
the EM waves emitted from 
the junctions are generated by the Chrenkov radiation due to the moving 
vortices \cite{Hechtfischer1},\cite{Hechtfischer2}. 
From our simuration results, we may suggest two possibilities for the 
origin of the broad-band emission. The first one is the Cherenkov-type 
radiation as claimed by Hechtfisher et al. Rough estimation for the 
present model indicates that the threshhold voltage for the Cherenkov 
radiation lies in region II, in which we really observe the broad-band 
spectrum in the low-frequency region as seen in Fig.3(II). It is noted that 
this result is consistent with simulation resutls by Hechtfischer et al
\cite{Hechtfischer2}. 
In this scenario, the step-like anomaly observed in the $I-V$ curve of Bi-2212 
may be identified with the boundary between the regions II and III in 
our simurations. 
On the other hand, it is also possible to interprete that 
the step coresponds to the boundary between the regions I and II, which leads to 
another possibility for the origin of the observed broad-band emmission. 
In this second scenario, it is understood that the observed EM field 
originates from the vortex state in region I. Since no transverse propagating 
Josephson plasma mode can be coupled with 
such the slow vortex motions in this region, one may interprete that the 
broad-band spectrum reflects chaotic EM field oscillations due to the 
non-regular motion of vortices. Here, note that 
step structure in IVC
at the boundary between the region I and II in simulations is 
the most remarkable structure and it is similar with experimental results 
\cite{Hechtfischer1}. 
To clarify the mechanism of the 
broad-band emission, we will need more detailed theoretical and experimental 
studies.

\section{Conclusion}

In conclusion, we performe direct large scale numerical simulations on 
Josephson vortex flow states in IJJ's
and find structural transitions of the moving flux-line lattice
accompanied by the step-like structures in the I-V characteristics and 
changes of power spectrums of AC electric field at the sample edge. 
The four typical flux flow states  
and the corresponding power spectra are identified and 
the in-phase 
superradiant flux flow state are found to exist 
stably in a wide region in the I-V characteristics.
The comparison between the simulation results and the eigenmode analysis 
for the transverse propagating Josephson plasma modes 
show that those step-like structures in the I-V characteristics
 are due to resonances between flux flow motions and  
transverse propagating Josephson plasma modes.  
Furthermore, the analysis predict that the   
superradiant state region enlarges 
as the superconducting and insulating layer thickness 
is decreased compared to the magnetic pentration depth in the ab-plane direction.
The collective Josephson vortex flow under the presence of 
many propagating modes includes very rich varieties of physics, while  
the stable superradiant flux flow state 
will be a key feature in making successfully  
submilimeter-wave generator employing IJJ's. 

         
The authors thank T.Yamashita, and S.Sakai for useful discussions 
and one of us ( M.M. ) also thank for T.Imamura and M.Itakura
for numerical simulation 
techniques and K. Asai and H. Kaburaki for  
their supports on the computer simulations in JAERI.



\end{multicols}\widetext

\end{document}